\documentclass[a4paper]{article}
\usepackage{setspace}
\usepackage{CJK}
\usepackage{amsfonts}
\usepackage[a4paper]{geometry}
\usepackage[pdftex]{graphicx}
\usepackage{epstopdf}
\usepackage{subfigure}
\usepackage[ansinew]{inputenc}
\usepackage{color}
\usepackage{cancel}
\usepackage{amsmath}
\usepackage{amssymb}
\usepackage{authblk}

\usepackage[a4paper=true,pagebackref=false]{hyperref}
\hypersetup{
   colorlinks = true,
   linkcolor = red,
   anchorcolor = red,
   citecolor = blue,
   filecolor = red,
   pagecolor = red,
   urlcolor = red}
   
   \doublespace
 \renewcommand{\thefootnote}{\alph{footnote}}

\begin{document}

\title{\Large \bf Near-field ablation threshold of cellular samples at mid-IR wavelengths.}

\author[1]{\rm Deepa Raghu}
\author[1,a]{\rm Joan A. Hoffmann}
\author[1,b]{\rm Benjamin Gamari }
\author[1,c]{\rm M.E. Reeves}
\affil[1]{\small{Department of Physics, The George Washington University, DC-20052, USA}}

\let\oldthefootnote\thefootnote
\renewcommand{\thefootnote}{\alph{footnote}}
\footnotetext[1]{Current address: The Johns Hopkins University Applied Physics Laboratory, MD-20723, USA}
\footnotetext[2]{Current Address, Department of Physics, The University of Massachusetts, MA-01003, USA}
\footnotetext[3]{Electronic mail: reevesme@gwu.edu}
\let\thefootnote\oldthefootnote

\maketitle
 



\begin{abstract}
We report the near-field ablation of material from cellulose acetate coverslips in water and myoblast cell samples in growth media, with a spot size as small as 1.5 $\mu$m under 3 $\mu$m wavelength radiation.  The power dependence of the ablation process has been studied and comparisons have been made to models of photomechanical and plasma-induced ablation. The ablation mechanism is mainly dependent on the acoustic relaxation time and optical properties of the materials. We find that for all near-field experiments, the ablation thresholds are very high, pointing to plasma-induced ablation as the dominant mechanism.
\end{abstract}
\underline{\hspace{\textwidth}}

Near-field scanning optical microscopy (NSOM)\cite{bib0} is a promising technique that overcomes the diffraction limit of conventional optical microscopy   
\cite{bib1,bib2} and by doing so has created a number of potential applications in biological imaging. The combination of NSOM for ablation with mass spectrometry is of particular interest to obtain detailed molecular information with spatial resolution better than that of the conventional optical spectrometry. A step in this direction, ultraviolet-NSOM-based mass spectrometry with a lateral resolution of 170 nm in ambient conditions, has demonstrated soft ablation capabilities \cite{bib3}.  An improvement would be ablating in the infrared rather than the UV regime so that the native water in a cell plays the role of the ablation matrix due to its strong absorption at 2940 nm \cite{irv,bib014}. By this approach, cells can be probed \emph{in-vivo} or \emph{in-vitro}, but in the far-field, the spatial resolution is limited by diffraction effects and by the quality of available optics to a spot size of about 50$\mu m$.  

There are in the literature a number of reports of ablation of conventional solids\cite{threshold} and organic molecules \cite{bib10,Zenobi08,Zenobi09,Meyer08}.  There are fewer in which the energy delivered to the sample is characterized well enough to measure ablation thresholds \cite{threshold,Zenobi09}. In these, the ablation thresholds are many orders of magnitude larger than would be expected for far-field approaches.

\begin{figure}[h]
\begin{center}
     \includegraphics[width=2.8in]{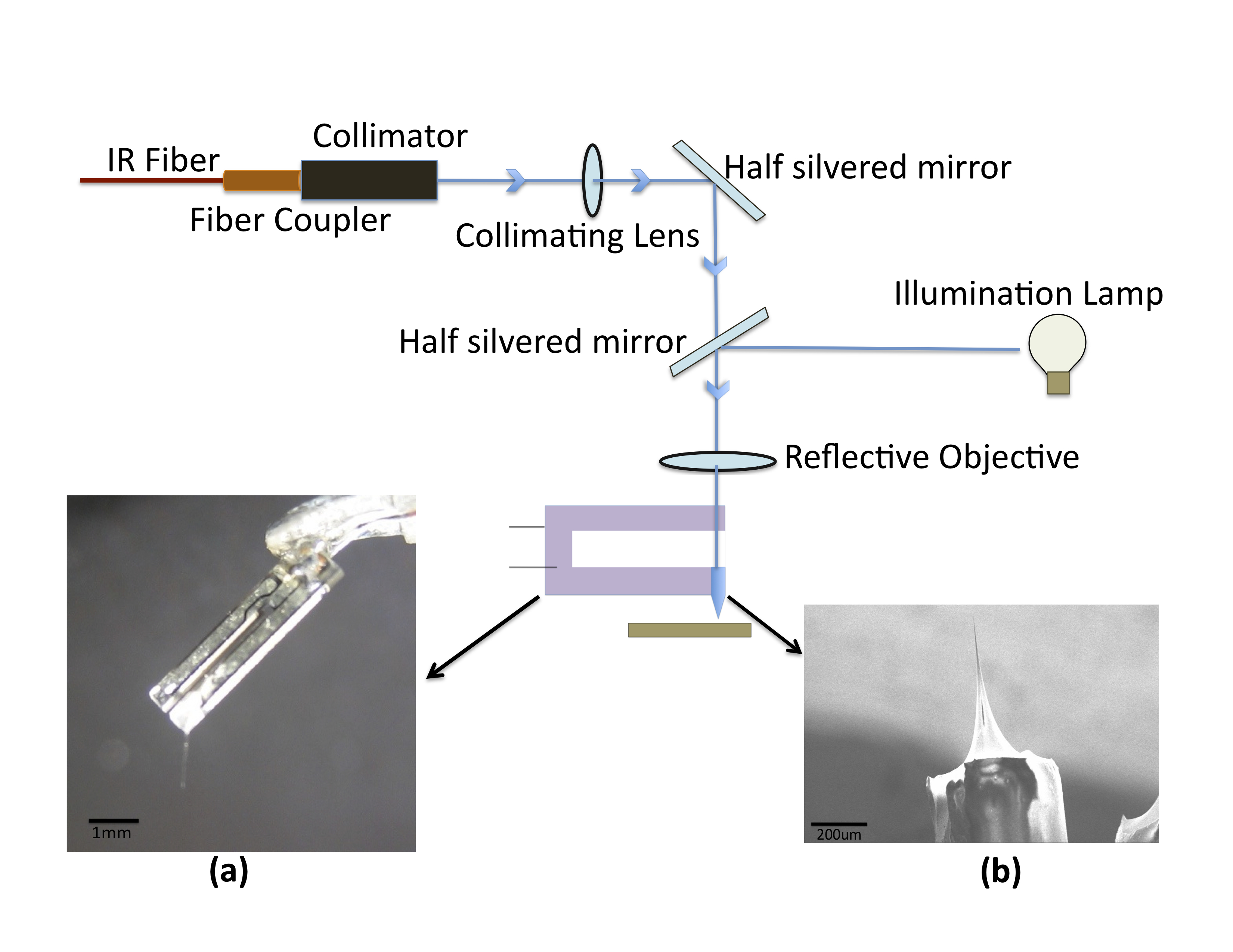}
     \caption{\label{scan1} Schematic diagram of NSOM setup, modified for IR ablation (a) IR fiber tip, attached to the tuning fork, scale bar:1 mm. (b) SEM image of an IR fiber tip etched using modified tube etching technique \cite{bib10,Hoffmann2011}, scale bar:200 $\mu$m.}
     \end{center}
  \end{figure}

We report the first integration of IR with near-field techniques applied to the ablation of live cells.  We obtain ablation features as small as 1.5 $\mu m$ under 3 $\mu m$ wavelength radiation, in hard and soft materials, and in air and water environments. The ablation threshold and fluence dependence of these processes are discussed here. Of particular interest is the successful ablation of cellular samples in growth media, which we describe in this paper.

The experiments described here are performed with an NSOM apparatus, modified for operation in the mid-IR region \cite{nanonics}. In all experiments, the mid-IR output of a Nd:YAG laser, coupled to a pumped-optical-parametric oscillator (OPO) (set to 2940 nm, 100 Hz, 5 ns pulse width) is used \cite{laser}. A schematic diagram of the experimental setup is shown in Fig.\ref{scan1}. A Schwarzschild objective is used to focus the laser beam and couple it to an NSOM tip, which is fabricated by tube etching  \cite{bib10,Hoffmann2011} from a short stub of germanium-based glass fiber \cite{irsystems}. AFM images of the sample, using normal force feedback, are obtained using the same tip. To calibrate the fluence through the tip,  laser power measurements are made using a PbSe photoconductive detector \cite{Gomella2011} positioned beneath the fiber tip in place of the sample.

\begin{figure}[h]
\begin{center}
     \includegraphics[width=3.5in]{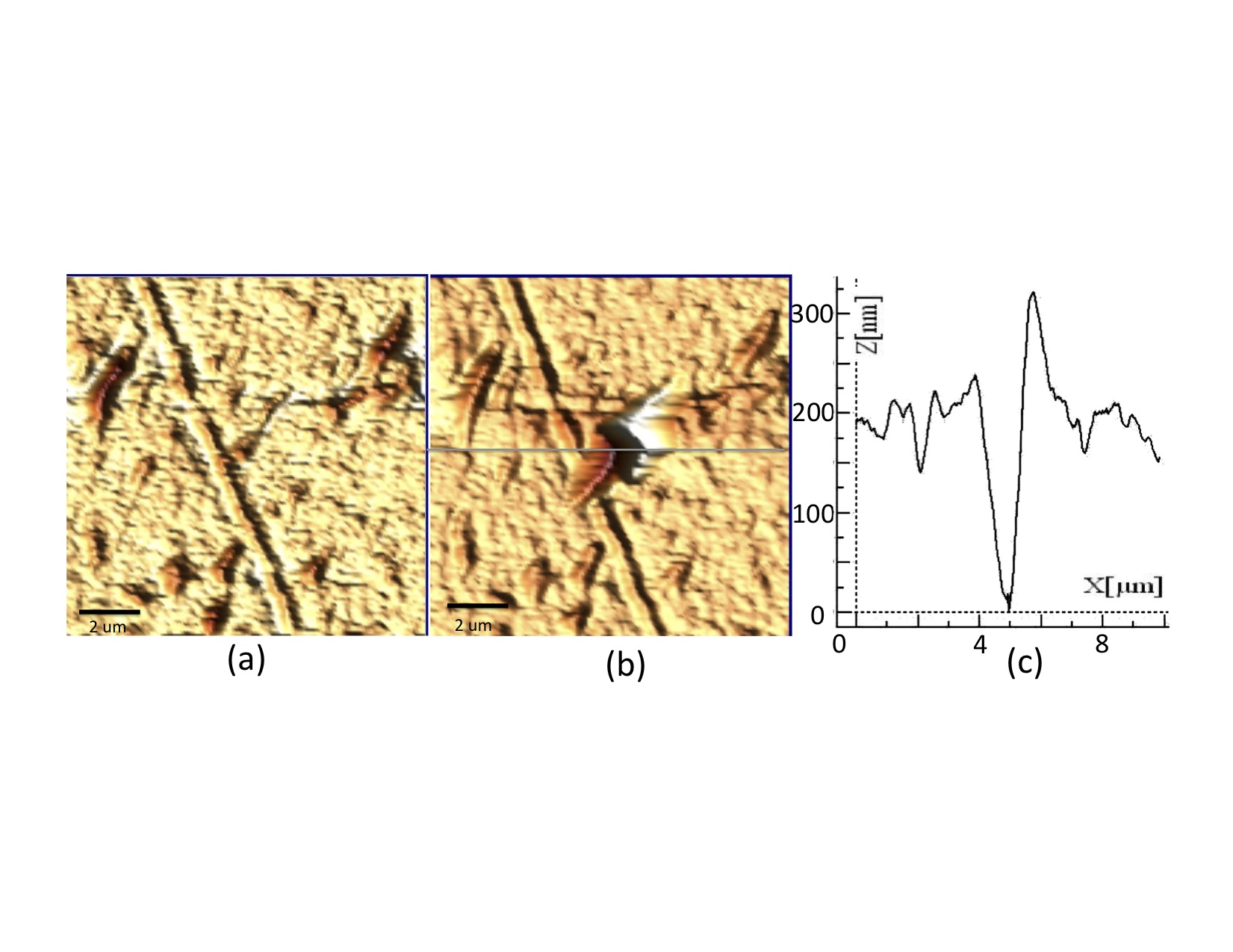}
     \vspace{-50pt}
     \caption{\label{scan} 3D Topographic image of a plastic cover slip in water (a) before and (b) after near-field ablation. (c) Profile of the crater, whose size is 1.25 $\mu$m, as measured by the full-width at half maximum (FWHM) of the profile. Scalebar: 2 $\mu$m}     
     \end{center}
     \end{figure}

To set a baseline for understanding the  ablation process, material from a cellulose acetate plastic coverslip is ablated in air in the far-field configuration. The ablation threshold for the sample is determined by conducting a series of single-spot ablations. In these, the laser beam is focused at normal incidence on the surface of the sample with a repetition rate of 100Hz. The sample is simultaneously observed under the microscope, and the ablation threshold is defined to be the fluence below which we do not observe visible permanent modification of the surface of the sample. In later experiments, a cellulose acetate plastic coverslip is ablated in air and in water by near-field techniques. Laser ablation in the water medium is achieved by immersing the cellulose acetate cover slip in water, 2 mm-deep,  and then coupling the laser beam into the tip, which is partially immersed in the water. The near-field ablation threshold is determined by comparing the AFM images for different laser powers. AFM images of the cellulose acetate coverslip before and after ablation in water are shown in Fig. \ref{scan}. 

\begin{figure}[h]
 \begin{center}
     \includegraphics[width=3.5in]{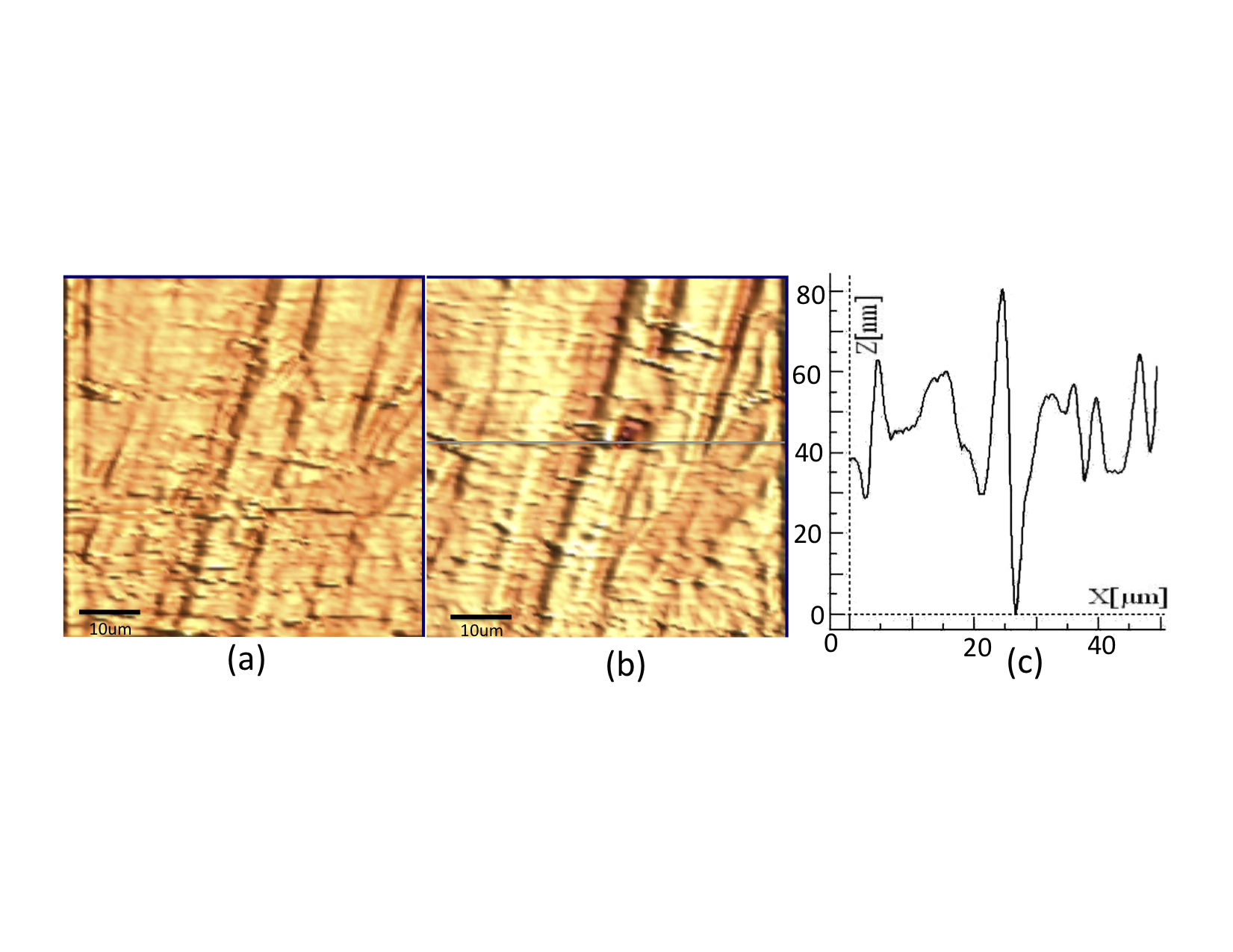}
      \vspace{-50pt}
 \caption{\label{myo2} Topographic image of myoblast cells in media (a) Before ablation. (b) After ablation. (c) Profile of the crater, whose size is 2.5 $\mu$m. Scalebar: 10 $\mu$m. }   
 \end{center} 
       \end{figure}

Myoblast cells are studied in specially prepared sample holders formed from a PDMS ring molded directly on a glass cover slip to hold the sample and surrounding media during the process of scanning and ablating. Cells from the C2C12 cell line are cultured directly in these molded wells.  Plated cells attach to the glass surface of the sample holders during the incubation period.  Topographic images of myoblast cells in media before and after ablation are shown in Fig. \ref{myo2}. Under the near-field laser illumination conditions of our experiment, it is clear from the AFM images that the damage is localized to a well defined region, approximately 2.5 $\mu$m in size.  From the profile images of the craters, we can see that there is some redeposition of the ablated material in the vicinity of the ablation crater, as is the case with the cellulose acetate material.

In our study, the ablation thresholds for a single material, cellulose acetate, varied greatly, from 0.22 J/cm$^2$ to 301 J/cm$^2$ as seen in Table I.  With such a big disparity between the near-field and the far-field ablation thresholds, the question naturally arises about whether the mechanism has differs.

In the far-field regime, the low thresholds are characteristic of a photomechanical ablation mechanism that starts with thermoelastic stresses arising from local expansion of the sample upon high-intensity laser heating\cite{stress1,diff}. The thermo-accoustic process is favored when the pulse width and spot size are suitable to locally contain the thermal and accoustic energy deposited by the beam \cite{threshold}. For example, the region of the sample ablated by each pulse is set by the thermal diffusion length $l_{th}$ \cite{threshold},
\begin{equation}
l_{th}=\sqrt{D t_p},
\end{equation}
where $t_p$ is the laser pulse width, and $D$ is the thermal diffusivity of the material.
For the cellulose acetate and myoblast  samples, the thermal diffusion length, $l_{th}$, is approximately 25 nm, small enough to minimize lateral thermal damage. Heating leads to local thermal expansion, and ablation occurs when conditions for stress or inertial confinement are achieved.  That is, for laser pulse durations shorter than a characteristic acoustic relaxation time\cite{confinement},
\begin{equation}
t_{ac}=\frac{1}{\alpha c_s}.
\end{equation}
Here $c_s$ is the speed of sound, and $\alpha$ is the optical absorption coefficient of the material. A theoretical model for the photomechanical mechanism has been developed by a number of research groups \cite{photomech1,mech1}, which indicates the possibility for thermo-accoustic ablation in our samples.  For cellulose acetate, the speed of sound in the sample is $c_s \approx 1340$ m/s and the absorption coefficient is $7481$ m$^{-1} $, to yield an acoustic relaxation time $t_{ac}$ of about  $100$ ns. This is much longer than the laser pulse width and satisfies the inertial confinement condition $t_p\ll t_{ac}$. 

For cellular samples in growth media, we expect the properties of water to determine the length and time scales for the ablation process.  The absorption coefficient of water is of the order of $1,200,000 $ m$^{-1}$ and the velocity of the sound is 1497 m/s, for which the acoustic relaxation time is equal to 0.5 ns. Comparing to our laser pulse width of 5 ns, the condition for the stress confinement, $t_{p} << t_{ac}$, is not satisfied, and photomechanical stress is not likely to be a factor, or at least will be reduced significantly.

\begin{table}[]
\begin{tabular}{|c|c|c|c|}
\hline
Material & spot size & ablation threshold   \\
&$\mu$m &  J/cm$^2$\\
\hline
cellulose acetate  (in air, ff) & 130 & 0.22\\

\hline
cellulose acetate  (in air, nf)  & 2.5 & 301 \\
\hline
cellulose acetate (in water, nf) & 2.5 &  135\\
\hline
Myoblast cells (in media, nf) & 2.5 & 61\\
\hline
\end{tabular}
\caption{Results of ablation threshold studies on cellulose acetate and myoblast cells in far-field (ff)  and near-field (nf) regime.  Each of these has been reproduced at least five times.  The laser wavelength is 2940 nm, and the pulse width is 5 ns }
\label{ourwork}
\end{table}

The measured thresholds for the near-field ablation for all cases reported here and for a number reported in the literature \cite{threshold,Zenobi09} are much higher than expected for photomechanical mechanism. This is not surprising for the ablation of cells but is so for the cellulose acetate. In both cases the measured ablation threshold points to a second, higher energy mechanism, plasma induced ablation \cite{ioni}. Here, the intense laser beam ionizes molecules in the sample, and subsequent collisions within the ablated plume and with ambient gas molecules lead to the formation of a hot dense plasma above the sample surface. The vapor plume then expands perpendicularly from the surface and is further ionized by incoming radiation. Consequently, the plasma absorbs more energy from the trailing part of the laser pulse by photoionization or inverse bremsstrahlung processes \cite{brem1,brem2}.

Near-field ablation alters the conditions for plasma-induced ablation.  The laser beam's path through the plasma is a narrow gap of only about 10 nm before reaching the sample, and hence the interaction with the plasma is weaker. Furthermore, the presence of a sharp probe in the vicinity of the sample leads to localization of the ionizing field and the probe itself physically blocks the expansion of the plume in the direction normal to the sample, leading to the movement of the plume laterally away from the ablated spot \cite{abl}. 

In water, the conditions for plasma-induced ablation are enhanced and we observe a decrease in the threshold fluence. This change is primarily due to the different optical properties of the water, namely the refractive index, which leads to an overall reflectivity decrease from $3.7\times10^{-2}$ to $4.1\times 10^{-3}$ upon immersion of the sample in water.  (The reflectivity is calculated in the usual way, R = $(n_1-n_2)^2/(n_1+n_2)^2$, here $n_2$ = 1.48, the refractive index of cellulose acetate.)  Overall, the absorptivity of water-sample system is larger than that of the air-sample system \cite{hao}. Also, in water, the plasma is confined and there is a delay in its expansion. Hence, the  induced pressure created by the laser sample interaction is much greater in water than in air and the plasma is compressed \cite{plasma}, an effect that is further enhanced by near-field confinement of the light. 

As has been observed by a number of groups \cite{threshold,Zenobi09} high fluences are required to achieve near-field ablation in organic \cite{bib3} and metallic samples \cite{abl}. Likewise, higher fluences lead to plasma induced ablation conditions for the myoblast cells. Due to the presence of water above and below the cell membrane, the induced plasma pressure in the myoblast cell samples is enhanced compared to that of the cellulose acetate sample. This effect is further amplified by the near-field confinement of the light. The reflectivity of myoblast cell samples, in growth media, is found to be $5 \times 10^{-4}$, assuming $n_{myoblast}=1.36$,\cite{curl2005} which helps to explain the smaller threshold fluence of myoblast cells compared to cellulose acetate.

In conclusion, as a first step towards \textit{in-vitro} mass spectral analysis of biological samples,  we have successfully demonstrated the ablation of hard and soft materials in liquids, and, in particular, of  biological samples in nutrient media. To understand the ablation mechanism, we have measured the ablation threshold of different types of samples. The mechanisms that are involved in the ablation process are mainly photomechanical and plasma induced ablation for samples in air. Since the acoustic relaxation time of water rich samples is of the order of the laser pulse width, the ablation mechanism in cellular samples is easily seen to be plasma induced rather than photomechanical. For cellulose acetate, the necessity for this large increase in the near-field ablation threshold in air is not as clear. However, for cells in media, their thermal and acoustic properties lead to conditions that favor the plasma-induced mechanism over the photomechanical one.

\section{Acknowledgments}
The authors would like to thank the W.M. Keck foundation for the financial support, Akos Vertes who provided the laser for our experiments, Infrared Fiber Systems, Silver Spring, MD, who provided the optical fibers for this study, William Rutkowsky for helping with the instrumentation, Andrew Gomella and Craig S Pelissier  for helping with the detector calibration process, Jyoti Jaiswal, Mary Ann Stepp and Gauri Tadvalkar for helping with the cell culture, and Alexander Jeremic for providing us the facility to culture cell samples.

\end{document}